\documentclass[prb, aps,showpacs, preprintnumbers, amsmath, amssymb, twocolumn, superscriptaddress]{revtex4-2}

\usepackage[a4paper, margin=15mm]{geometry}
\usepackage [latin1]{inputenc}
\usepackage{graphicx,pbox,lineno}
\usepackage[dvipsnames]{xcolor}
\usepackage[normalem]{ulem}
\allowdisplaybreaks
\usepackage[colorlinks, citecolor={blue!50!black}, urlcolor={blue!50!black}, linkcolor={red!50!black}]{hyperref}
\usepackage[caption=false]{subfig}
\usepackage{units}
\usepackage{mathtools}
\usepackage{amsmath}
\usepackage{amssymb}
\usepackage{bbold}

\usepackage{gensymb}
\def\be{\begin{equation}}
\def\ee{\end{equation}}
\def\bea{\begin{eqnarray}}
\def\eea{\end{eqnarray}}

\begin{document}


\title{Dynamic Spin Fluctuations in the Frustrated Spin Chain Compound Li$_3$Cu$_2$SbO$_6$}

\author{A. Bhattacharyya}
\email{amitava.bhattacharyya@rkmvu.ac.in}
\affiliation{Department of Physics, Ramakrishna Mission Vivekananda Educational and Research Institute, Belur Math, Howrah 711202, West Bengal, India} 
\author{T. K. Bhowmik}
\affiliation{Department of Physics, Bose Institute, 93/1, Acharya Prafulla Chandra Road, Kolkata, 700009, India}
\author{D. T. Adroja} 
\email{devashibhai.adroja@stfc.ac.uk}
\affiliation{ISIS Facility, Rutherford Appleton Laboratory, Chilton, Didcot Oxon, OX11 0QX, United Kingdom} 
\affiliation{Highly Correlated Matter Research Group, Physics Department, University of Johannesburg, PO Box 524, Auckland Park 2006, South Africa}
\author{B. Rahaman}
\affiliation{Department of Physics, Aliah University, Newtown, Kolkata 700156, India}
\author{S. Kar}
\affiliation{Department of Physics, A.K.P.C. Mahavidyalaya, Hooghly - 712611, India}
\author{S. Das}
\author{T. Saha-Dasgupta}
\email{tanusri@bose.res.in}
\affiliation{S.N. Bose National Centre for Basic Sciences, Salt Lake, Kolkata 700106 India}
\author{P. K. Biswas} 
\affiliation{ISIS Facility, Rutherford Appleton Laboratory, Chilton, Didcot Oxon, OX11 0QX, United Kingdom}
\author{T. P. Sinha}
\affiliation{Department of Physics, Bose Institute, 93/1, Acharya Prafulla Chandra Road, Kolkata, 700009, India}
\author{R. A. Ewings}
\affiliation{ISIS Facility, Rutherford Appleton Laboratory, Chilton, Didcot Oxon, OX11 0QX, United Kingdom} 
\author{D. D. Khalyavin} 
\affiliation{ISIS Facility, Rutherford Appleton Laboratory, Chilton, Didcot Oxon, OX11 0QX, United Kingdom} 
\author{A. M. Strydom}
\affiliation{Highly Correlated Matter Research Group, Physics Department, University of Johannesburg, PO Box 524, Auckland Park 2006, South Africa}
\affiliation{Max Planck Institute for Chemical Physics of Solids, Nothnitzerstr. 40, D-01187 Dresden, Germany}

\begin{abstract}

We report the signatures of dynamic spin fluctuations in the layered honeycomb Li$_3$Cu$_2$SbO$_6$ compound, with a 3$d$ S = 1/2 $d^9$ Cu$^{2+}$ configuration, through muon spin rotation and relaxation ($\mu$SR) and neutron scattering studies. Our zero-field (ZF) and longitudinal-field (LF)-$\mu$SR results demonstrate the slowing down of the Cu$^{2+}$ spin fluctuations below 4.0 K. The saturation of the ZF relaxation rate at low temperature, together with its weak dependence on the longitudinal field between 0 and 3.2 kG, indicates the presence of dynamic spin fluctuations persisting even at 80 mK without static order. Neutron scattering study reveals the gaped magnetic excitations with three modes at 7.7, 13.5 and 33 meV.  Our DFT calculations reveal that the next nearest neighbors (NNN) AFM exchange ($J_{AFM}$ = 31 meV) is stronger than the NN FM exchange ($J_{FM}$ = -21 meV) indicating the importance of the orbital degrees of freedom. Our results suggest that the physics of Li$_3$Cu$_2$SbO$_6$ can be explained by an alternating AFM chain rather than the honeycomb lattice.

\end{abstract}

\date{\today} 


\maketitle


\noindent {Low dimensional Honeycomb layered oxide materials that consist of alkali metal atoms sandwiched between slabs of transition metal and chalcogen or pnictogen atoms arranged in a honeycomb fashion are of great interest at present because these materials play host to fascinating symmetry-protected topological phases and are crucial for next-generation cathode materials for rechargeable batteries~\cite{Li2020battery}. Emergent properties they exhibit are the Haldane gap, fractionalization of spin degrees of freedom, and a topological quantum spin liquid (QSL) state with exotic quasiparticles for honeycomb lattice in which the spins fractionalize into emergent quasiparticle-Majorana fermions~\cite{savary2016quantum, zhou2017quantum, banerjee2016proximate, balents2010spin, yan2011spin, helton2007spin}.  Experimental progress concerning QSL states in realistic materials, including organic anisotropic triangular-lattice Mott insulators [$\kappa$-(ET)$_2$Cu$_2$(CN)$_3$~\cite{Yamashita2008} and EtMe$_3$Sb[Pd(dmit)$_2$]$_2$]~\cite{Yamashita2011}, the Kagome-lattice system [ZnCu$_3$(OH)$_6$Cl$_2$]~\cite{mendels2010quantum}, and the three-dimensional hyperkagome lattice system Na$_4$Ir$_3$O$_8$~\cite{shockley2015frozen}. Despite the large magnetic exchange $J\approx$ 250 K observed in these systems~\cite{Yamashita2008,Yamashita2011,mendels2010quantum,shockley2015frozen}, there is no experimental indication of long-range magnetic ordering down to a temperature of $\sim$30 mK. Candidate materials for QSL are 5$d$ and 4$d$ transition metal compounds with the $d^5$-electron configuration such as iridates $\alpha$-Na$_2$IrO$_3$, $\alpha$-Li$_2$IrO$_3$, H$_3$LiIr$_2$O$_6$, Ag$_3$LiIr$_2$O$_6$, and ruthenium-based $\alpha$-RuCl$_3$~\cite{HwanChun2015Na2IrO3,Katukuri2016Li2IrO3,Kitagawa2018H3LiIrO3,Bahrami2019Ag3LiIr2O6,Takagi2019QSL}. Iridates materials crystallize in an alternating 2D layered structure in which IrO$_6$ octahedra form a honeycomb network by sharing the three orthogonal edges of an octahedron with 90$^\circ$ Ir-O-Ir bonds. The orthogonal anisotropies of the three nearest-neighbor bonds of each spin conflict with each other, leading to frustration. QSL state has been seen for H$_3$LiIr$_2$O$_6$, other candidates of LiIr$_2$ type were magnetically ordered, due to Heisenberg interactions caused by $d-d$ exchange coupling which compete with Kitaev-type interactions, and supporting a magnetically ordered state~\cite{Takagi2019QSL}.}


\noindent {In the case of the 3$d$ $d^7$ quasi-two-dimensional (2D) honeycomb lattice A$_3$A$^{\prime}_2$BO$_6$ (A = Li, Na; A$^{\prime}$ = Co, Ni; B = Sb, Te;) depending upon the anisotropy and frustration triggered by the competition between AFM and ferromagnetic (FM) exchange interactions, numerous forms of unusual ordering are found such as FM, AFM, zig-zag AFM and stripe order AFM, which is anticipated to have the same origin as in the $d^{5+}$ materials, that is due to Kitaev-type interactions~\cite{WONG2016Na3Co2SbO6,Wang2017Na2Ni2TeO6,Kurbakov2017Li3Ni2SbO6,Scheie2019Na3Ni2BiO6}. Turning to the title material of this paper, the $d^9$ Li$_3$Cu$_2$SbO$_6$ compound crystallizes in a distorted honeycomb lattice with edge-sharing CuO$_6$ octahedra with space group $C2/m$~\cite{koo2016static}. The bond geometry of the Cu-O-Cu bond angle resembling $\approx$ 90$^\circ$ puts this material close to a QSL like state~\cite{miura2008magnetic, koo2016static}. The frustration index defined as $f$ = |$\theta/T_{N,f}$|, where $\theta$ is the Weiss temperature, and $T_{N,f}$ is either the Neel temperature or the spin freezing temperature. The lack of a magnetic transition down to 50 mK in Li$_3$Cu$_2$SbO$_6$ indicates that it is highly frustrated. Here we present the ground state spin dynamics of Li$_3$Cu$_2$SbO$_6$ through a muon spin relaxation study. Notably, below $\sim$ 4.0 K, a novel and unusual spin state appears, which does not reveal any magnetic ordering down to 50 mK. Neither the oscillations in the time dependent asymmetry nor 2/3 drop in the initial asymmetry are observed in the ZF-spectra down to 80 mK conforming the absence of long range ordering. The absence of magnetic peaks is confirmed by neutron diffraction data at 50 mK, thus, ruling out any long-range magnetic ordering. The muon spin relaxation rate measured in ZF exhibits a plateau below 1.0 K. These observations suggest that a QSL like ground state is formed in Li$_3$Cu$_2$SbO$_6$. Furthermore, our inelastic neutron study reveals the presence of gapped magnetic excitations at 7 K, which can be explained using the exchange parameters estimated from our DFT calculation.}


\begin{figure}[t]
\centering
\includegraphics[scale=0.25]{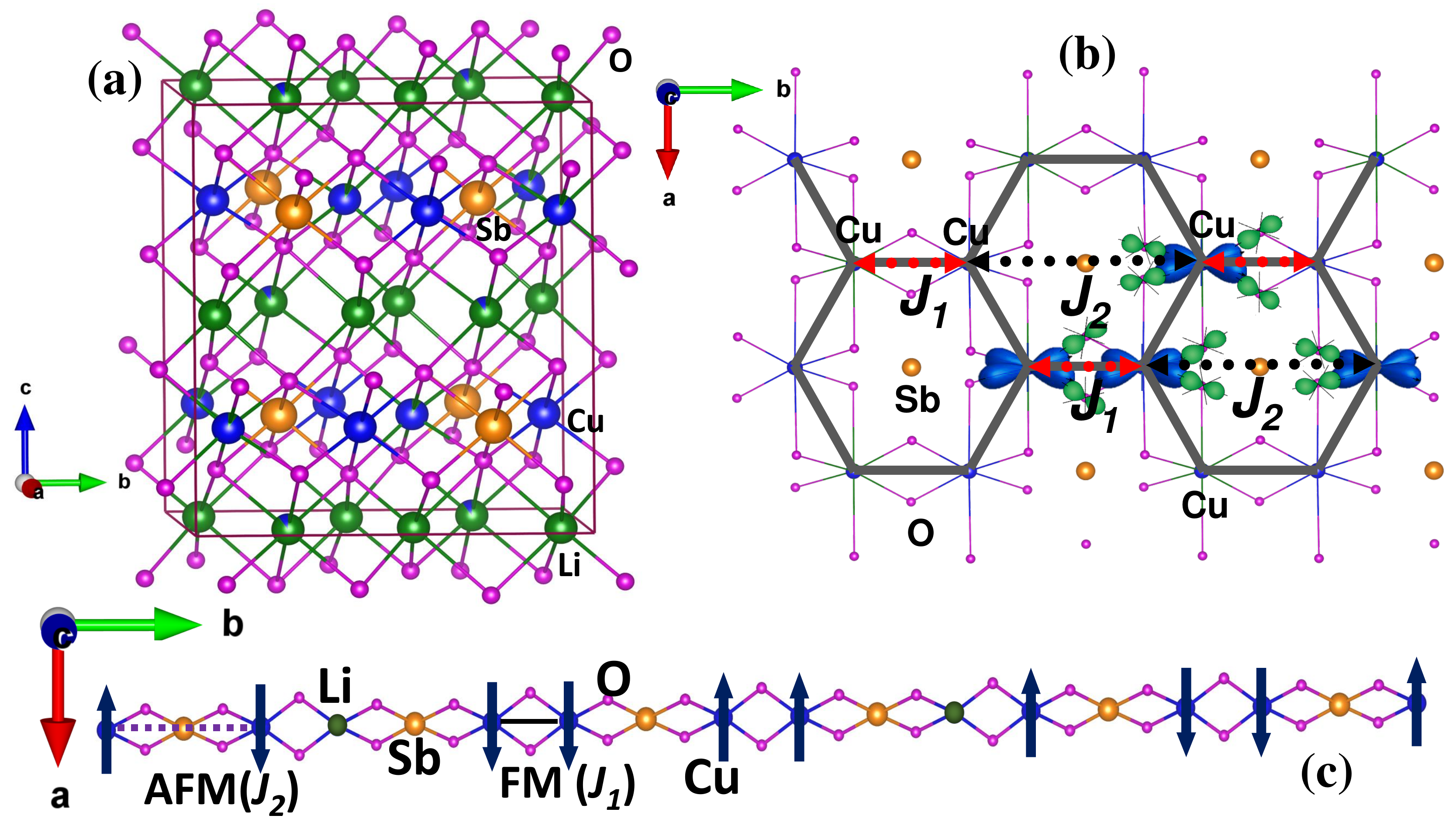}
\caption{\textcolor{black}{(a) The monoclinic unit cell representation of Li$_3$Cu$_2$SbO$_6$, (b) Sketch of the ab-plane of Li$_3$Cu$_2$SbO$_6$ honeycomb structure (the grey hexagon illustrates the underlying honeycomb structure), Cu-O bond length and Cu-O-Cu bond angles are given in figure. (c) The FM-AFM chain formed by Cu ions along the $b$-axis in the ab-plane whereas the Li-ion acts as the non-magnetic defects and produces the fragmented Cu-spin chains.}}
\label{heatcapacity}
\end{figure}


\noindent {\bf Sample preparation and characterization.} Single phase polycrystalline Li$_3$Cu$_2$SbO$_6$ sample was prepared employing a solid state reaction method. Li$_2$CO$_3$ (BDH, 99\%), Sb$_2$O$_3$ (Aldrich, 99.99\%) and CuO (Cerac, 99.999\%) powder was thoroughly ground and pressed into a pellet. The pellet was placed in an alumina boat and heated to 900 $^{\circ}$C. It was cooled down to 600 $^{\circ}$C in 100 h in air finally, the furnace was turned off. Magnetization measurements were performed using superconducting MPMS-XL7AC. The specific heat was measured using PPMS with a $^3$He/$^4$He dilution refrigerator. High-resolution neutron powder diffraction, $\mu$SR and inelastic neutron scattering data were collected at the ISIS Pulsed Neutron and Muon Source of Rutherford Appleton Laboratory, U.K.  

\noindent {\bf Crystal Structure and thermodynamics properties:} The crystal structure was refined by the Rietveld method using neutron diffraction (ND) data collected at 50 mK and 10 K, as presented in supplementary figure 1(a-b). There is no difference between these two ND spectra. No new peaks are detected down to 50 mK, which is $\sim$ 4 orders of magnitude lower than the $J_{AFM}$ value of 31 meV (359 K), indicating the absence of long-range magnetic ordering. The distorted honeycomb phase was indexed, which crystallizes in the monoclinic space group $C2/m$ as shown in Figure 1(a-c), isostructural with other compounds in this series~\cite{Kurbakov2017Li3Ni2SbO6,vinod2013c2m,stratan2019c2m}. The $\chi(T)$ data as shown in supplementary figure 2(a) does not show any feature of long-range magnetic order down to 1.8 K in agreement with ND data. The linearity of inverse $\chi(T)$ data at high $T >$ 200 K (I) and low $T <$ 30 K (II) temperatures suggest two different regions of Curie-Weiss (CW) like behavior. The analysis yields that the number of S = 1/2 Cu$^{2+}$ spins effective at high $T$ is about 1.0/f.u.
The difference of $\chi^{exp}-\chi^{CW^I}$ presented in the inset of supplementary figure 2(a), shows thermally activated behavior. The peak appears near 80 K. The $\chi(T)$ data well accounted with our calculated FM-AFM exchange interactions with $J_{FM}$ = -21 meV and $J_{AFM}$ = 31 meV~(see DFT calculation). Absence of $\lambda$-type anomaly in heat capacity data rules out the possibility of any long-range ordering, which supports the susceptibility and neutron diffraction data as shown in supplementary figure 2(b). A Schottky-like feature in the heat capacity is observed around $\approx$ 2 K, as shown in the inset of supplementary figure 2(b). Similar anomalies observed in herbertsmithite KCu$_6$AlBiO$_4$(SO$_4$)$_5$Cl, this behaviour was attributed to short-range spin correlations on the interlayer sites~\cite{Fujihala2020,FUKAYA2000185}. It is interesting to note that the magnetic entropy reaches a value of 0.5Rln2 near 16 K, suggesting short range ordering (supplementary figure 2c). \\

\noindent {\bf ZF-$\mu$SR: Evidence for short-range correlations.} 
No indication of long-range magnetic ordering has been found down to 50 mK through neutron diffraction, magnetic susceptibility, or specific heat measurements which motivated us to measure the zero-field (ZF) and longitudinal field (LF) muon spin relaxation ($\mu$SR) of Li$_3$Cu$_2$SbO$_6$ down to 80 mK. ZF/LF-$\mu$SR is an exceptional tool to probe static and dynamic magnetic fluctuations or quantum magnetism and often employed to unravel the enigmatic QSL state. Figure 2(a) represents the ZF-$\mu$SR data at different temperatures, which consists of the local responses of muons embedded at various stopping sites of Li$_3$Cu$_2$SbO$_6$. Our important observations from the ZF-$\mu$SR data are the following (a) the lack of oscillations in muon spectra as usually seen for magnetically ordered systems and (b) no loss of the initial asymmetry at time $t$ = 0 down to 80 mK. These observations strongly suggest the absence of static magnetism in Li$_3$Cu$_2$SbO$_6$ and signify slowing down of the spin dynamics. We have used several relaxation functions to fit our ZF, and LF-$\mu$SR data, starting from (a) simple exponential decay (b) stretched exponential, and (c) Umerao spin-glass. The best fit is obtained using a stretched exponential relaxation plus decaying exponential function with a constant background term, $G_z(t) = A_1\exp[-(\lambda_{ZF_1} t)^{\beta_{ZF}}]+A_2\exp(-\lambda_{ZF_2} t)+ A_{bg}$, here, the first two terms reflect the contribution of the muons that stop within the sample, and the third term accounts for those muons that stop within the sample holder. The slow and fast exponential decays $\lambda_{ZF_1}$ and $\lambda_{ZF_2}$ respectively, represents a two-component electronic spin contribution to the muon depolarization. A$_1\sim$ 21\% and A$_2\sim$ 4\% are the initial asymmetry values, $\lambda_{ZF_1}$/$\lambda_{ZF_2}$ are the muon spin relaxation rate, and $\beta_{ZF}$ is the stretching exponent. This approach has also been used in the analysis of the QSL like candidates SrCr$_2$Ga$_8$O$_{19}$~\cite{Uemura1994musr} and Sr$_2$Cu(Te$_{0.5}$W$_{0.5}$)O$_{6}$~\cite{Mustonen2018musr}. The coefficient A$_ {bg}\sim$ 0.5\% is a background constant representing muons that missed the sample. The solid curves show the fits of ZF-$\mu$SR data in Figure 2(a). Figure 2(b) shows the temperature dependence of the spin depolarization rate, $\lambda_{ZF_1}$,$\lambda_{ZF_2}$, and the stretching component, $\beta_{ZF}$ (supplementary figure 3a-b), respectively. 

\begin{figure}[t]
\centering
\includegraphics[width=0.46\linewidth, height=0.6\linewidth]{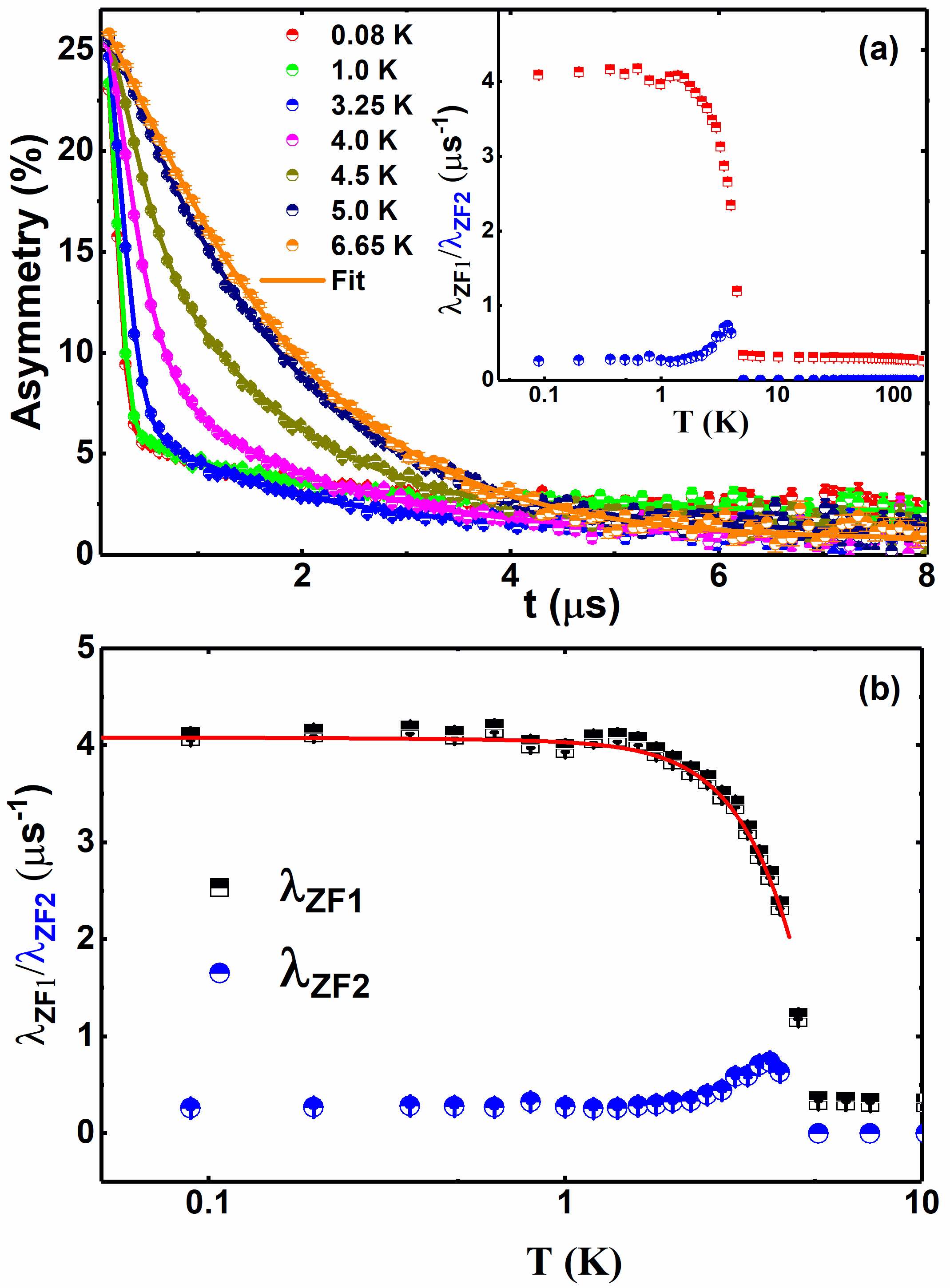}
\includegraphics[width=0.46\linewidth, height=0.6\linewidth]{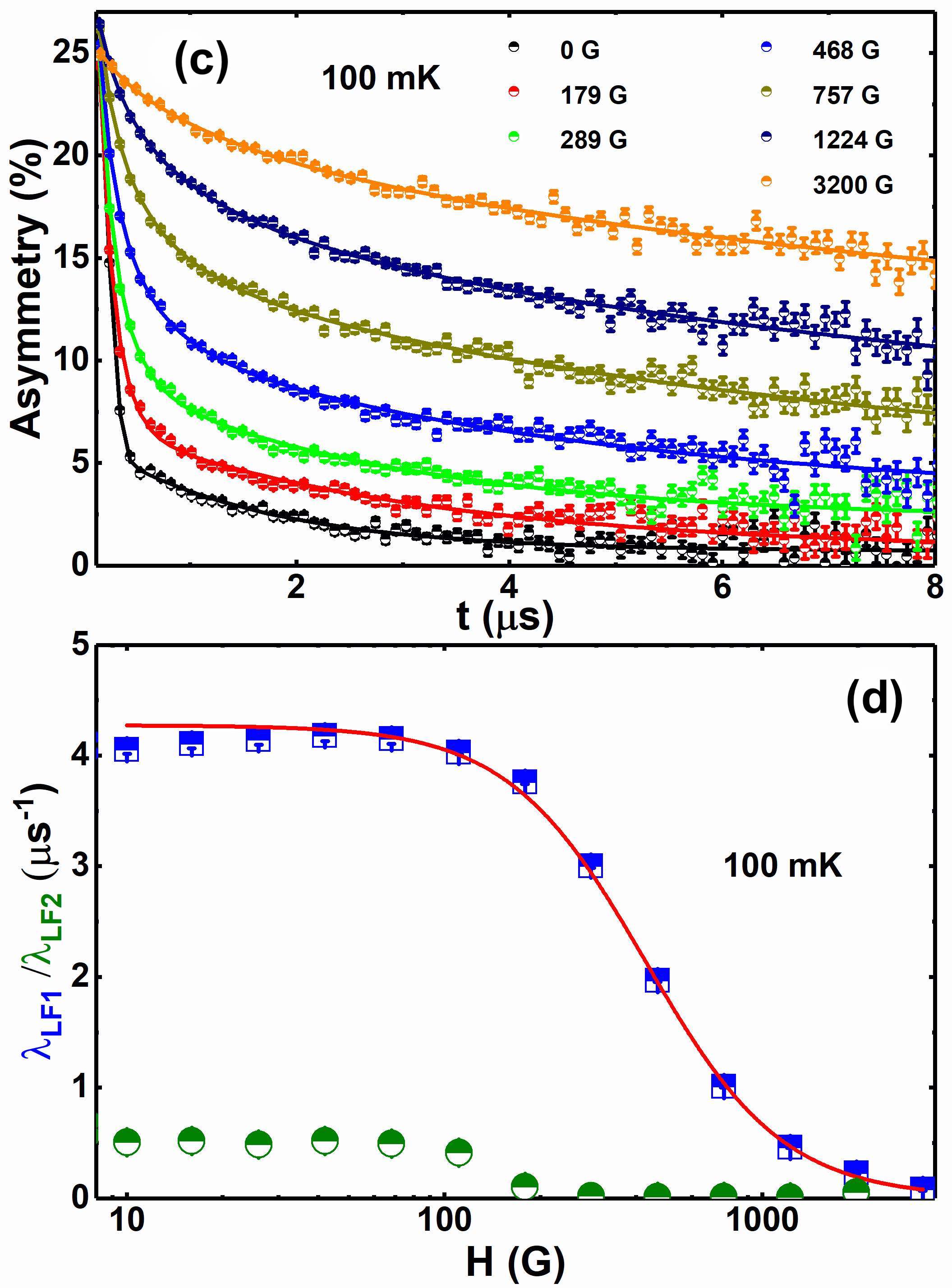}
\caption{(a) Time-dependent muon spin polarization of Li$_3$Cu$_2$SbO$_6$ at selected temperatures measured in zero field. Solid lines are fits to the data using equation (1). This equation consists of two relaxation processes, $\lambda_{ZF_1}$ and $\lambda_{ZF_2}$ are the fast and slow relaxation rates, respectively. The two relaxation processes were found to have the weighting ratio 5:1. This is consistent with the ratio of the two muon sites expected from the crystal structure. (b) Temperature dependence of $\lambda_{ZF_1}$ and $\lambda_{ZF_2}$. (c) Time-dependent muon spin polarization of Li$_3$Cu$_2$SbO$_6$ at selected longitudinal applied fields at 100 mK. Solid lines are fits to the data using equation (1). (d) LF dependence of the muon spin relaxation rate, at 100 mK. The line represents the fit to the data using equation (2).}
\label{zfmusr}
\end{figure}

\noindent The important result shown in Figure 2(b) is that there is no static magnetism in Li$_3$Cu$_2$SbO$_6$ on cooling to at least 80 mK which agrees with the neutron diffraction, susceptibility, and heat capacity data as presented above. Therefore the magnetic fluctuations remain dynamic down to the lowest temperature of measurements, an indispensable criterion for a system to undergo any transition to a spin freezing state. The rise of $\lambda_{ZF_1}$ below 4 K designates the slowing down of Cu$^{2+}$ spin fluctuations as a consequence of short-range interactions. It is interesting to note that below 1 K, the $\lambda_{ZF}$ shows a temperature-independent plateau-like behavior (see Figure 2(b)), which signifies the presence of dynamic spin fluctuations. The plateau-like behavior in $\lambda_{ZF}$ vs. $T$ data has also been seen in other QSL candidates~\cite{gomil2016musr,zorko2008musr}. The stretching exponent $\beta_{ZF}$ (supplementary figure 3a) approaches to 1 near 2 K, which suggests fast fluctuating local internal fields. Below 2 K, $\beta_{ZF}$ increases with decreasing temperature and attains a maximum value of $\sim$ 2 at 80 mK, suggesting Gaussian field distributions produced by magnetic exchange interactions of nearest neighbor spins. This type of behavior of $\beta_{ZF}$ designates the slowing down of magnetic fluctuations at low temperatures. Similar $\beta_{ZF}$ values are observed in the QSL candidates SrCr$_2$Ga$_8$O$_{19}$~\cite{Uemura1994musr} and Sr$_2$Cu(Te$_{0.5}$W$_{0.5}$)O$_{6}$~\cite{Mustonen2018musr}.\\

\noindent {\bf LF-$\mu$SR: Probing the spin-spin correlations.} To understand the nature of dynamic spin fluctuations, we further examine the LF-$\mu$SR data at different fields from 1 mT up to 320 mT measured at 100 mK, as shown in Figure 2(c). The size of internal field distributions is estimated as $\Delta$ = 1 mT, where $\gamma_\mu$ = 135.5$\times$2$\pi$s$^{-1}$$\mu$T$^{-1}$ is the gyromagnetic ratio for muons. The absence of saturation of LF-$\mu$SR spectra at 320 mT infers that the plateau-like behavior is not associated with static magnetic fluctuations, as seen for magnetically ordered systems. Hence it must be linked with dynamic fluctuations as observed for QSL like systems~\cite{sarkar2019musr}. The field dependence of $\lambda_{LF}$ is shown in Figure 2(d), the observed plateau-like behavior suggests the slowing down of spin fluctuations. This could be associated with spin liquid like behavior similar to $\mu$SR observations reported for Ce$_2$Sn$_2$O$_7$~\cite{Gao2019broadpeakCm} and SrDy$_2$O$_4$~\cite{petrenko2017musr}. We have fitted $\lambda_{LF}(H)$ using Redfield's equation, $\lambda_{LF} = \lambda_{LF_0}+\frac{2\gamma^2_{\mu}\Delta^2\tau_C}{1+\gamma^2_{\mu}H^2\tau^2_C}$, where $\Delta$ represents the amplitude of the internal field distribution, and the relaxation timescale of spin fluctuations is $\tau_{C}$. The fit to the $\lambda_{LF}(H)$ data is given by the solid red line in Figure 2(d). The calculated parameters are $\lambda_{LF}(0)$ = 0.003 $\mu s^{-1}$, $\Delta$ = 1 mT, and $\tau_C$ = 2.7$\times$ 10$^{-8}$ s.\\

\begin{figure}[t]
\centering
\includegraphics[width=0.55\linewidth, height=0.6\linewidth]{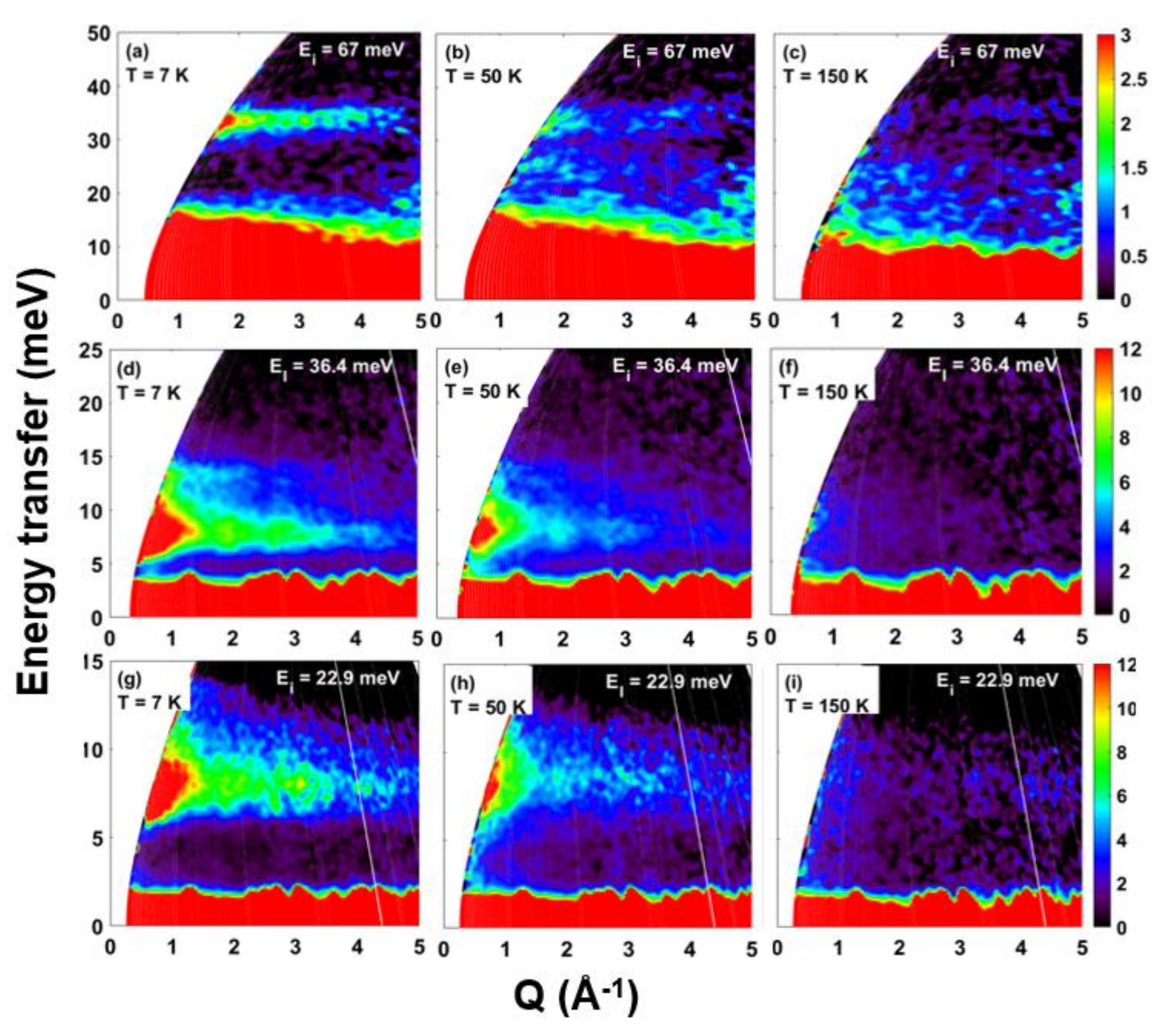}\hfill
\includegraphics[width=0.45\linewidth, height=0.6\linewidth]{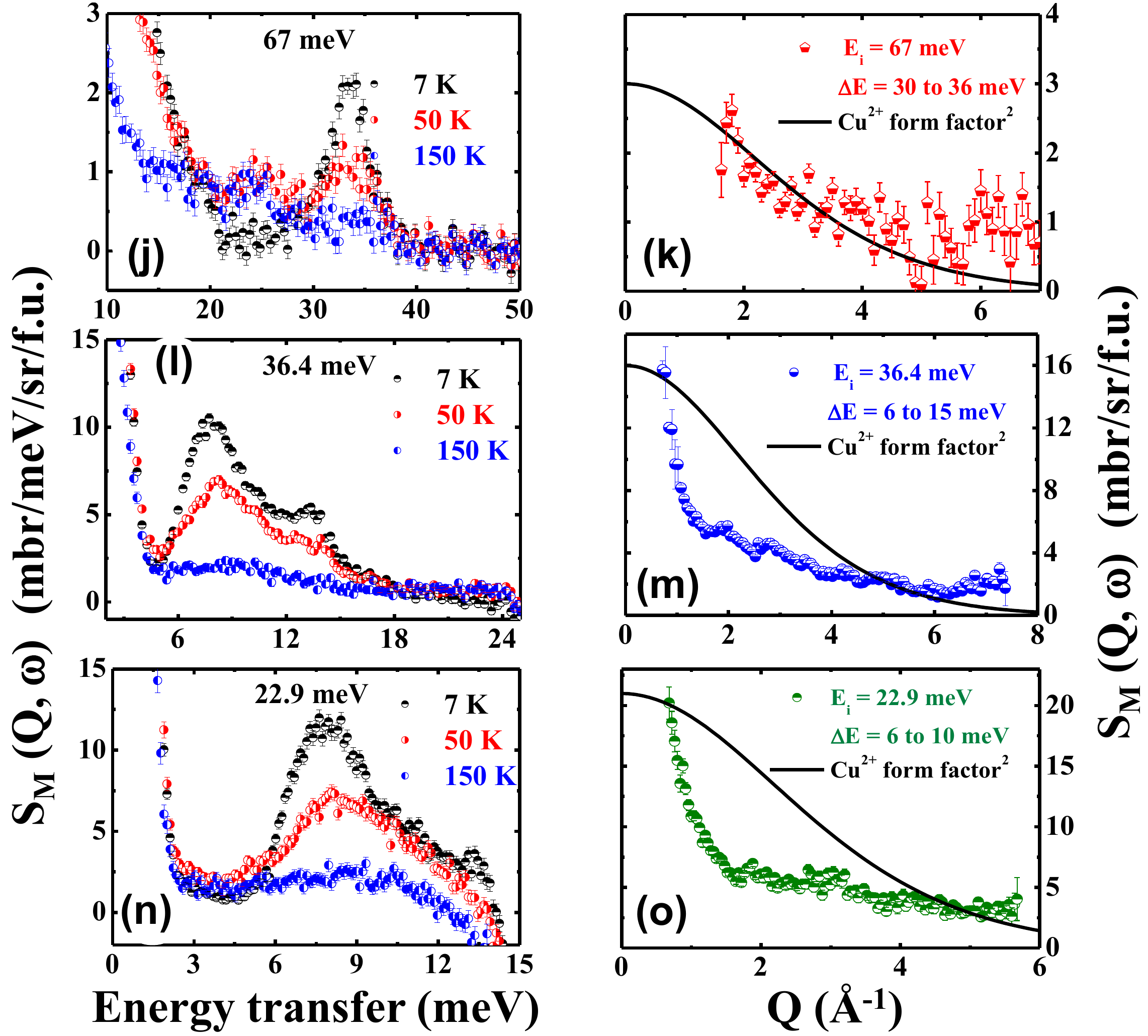}
\caption{(Left panel, a-i) shows the estimated magnetic scattering of Li$_3$Cu$_2$SbO$_6$ at 7, 50 and 150 K obtained after subtracting phonon scattering. (a-c) is for $E_i$ = 67 meV, (d-f) is for $E_i$ = 36.4 meV and (g-i) is for $E_i$ = 22.9 meV. The strong scattering near zero energy transfer is mainly due to the incoherent background and elastic scattering. (Right panel, j-o) The scattering angle integrated (2$\theta$ = 8$^\circ$ to 35$^\circ$) 1D-energy cuts of the magnetic intensity, from the color maps, plotted as Intensity vs Energy transfer at 7, 50 and 150 K (j) for $E_i$ = 67 meV, (l) for $E_i$ = 36.4 meV and (n) for $E_i$ = 22.9 meV. Note that the slightly negative signal at larger energy transfer, especially in 22.9 meV, is an artifact of the phonon background subtraction due to frame-overlap in multi-$E_i$ measurements. (Right panel k-m-o) The energy integrated momentum dependent ($Q$) magnetic intensity of three magnetic excitations at 7 K. The solid line shows the normalized magnetic form factor squares ($F^2(Q)$) of Cu$^{2+}$ ion.}
\label{zfmusr}
\end{figure} 

\noindent {\bf Inelastic Neutron Scattering: Magnetic excitations.} The results of INS study for incident energies $E_i$ = 67, 36.4, and 22.9 meV are presented in Figure 3(a-i) as color maps of the scattering intensity, energy transfer versus momentum transfer ($Q$) at 7 K, 50 K, and 150 K. The data have been corrected with the phonon scattering using the measured data at 300 K and using the Bose factor. Further, the one-dimensional, scattering angle integrated (or $Q$-integrated), energy cuts from the color intensity maps at the lower scattering angles, where the magnetic scattering is dominated as it follows the magnetic form factor squares, $F^2(Q)$, of Cu$^{2+}$ ion, and higher scattering angles, where the phonon scattering is dominated as the phonon scattering increases as $Q^2$, are plotted in supplementary figure 4. Figures 3(a, d, g) show that at 7 K, we have three magnetic excitations near 33 meV, 13.5, and 7.7 meV. With the increasing temperature at 50 K, the intensity of all three excitations decreases (see Figure 3). It is interesting to note that at 50 K, no apparent change in the phonon intensity at high angles (see supplementary figure 4) has been seen compared with 7 K in 67 meV data. In comparison, at a low angle, a weak increase in the intensity near 25 meV has been observed at 50 K compared to 7 K. Further, at 150 K, the intensity of all three modes (7.7 meV, 13.5 meV, and 33 meV) is reduced considerably. Upon closer inspection, the color plots of 150 K in Figure 3 indicate that scattering has moved to lower $Q$ and lower energy at 150 K, which might suggest that ferromagnetic like short-range correlations exist in the high-temperature range. This has also been supported by the data of 36.4 meV and 22.9 meV at 150 K. To check whether the energy of three observed magnetic excitations exhibits any dispersion (i.e., $Q$ dependence) at 5 K, we made 1D cuts at various $Q$-positions (not shown here) and found that the energy of all the magnetic excitations are near $Q$-independent.


\begin{figure}
\includegraphics[width=\linewidth]{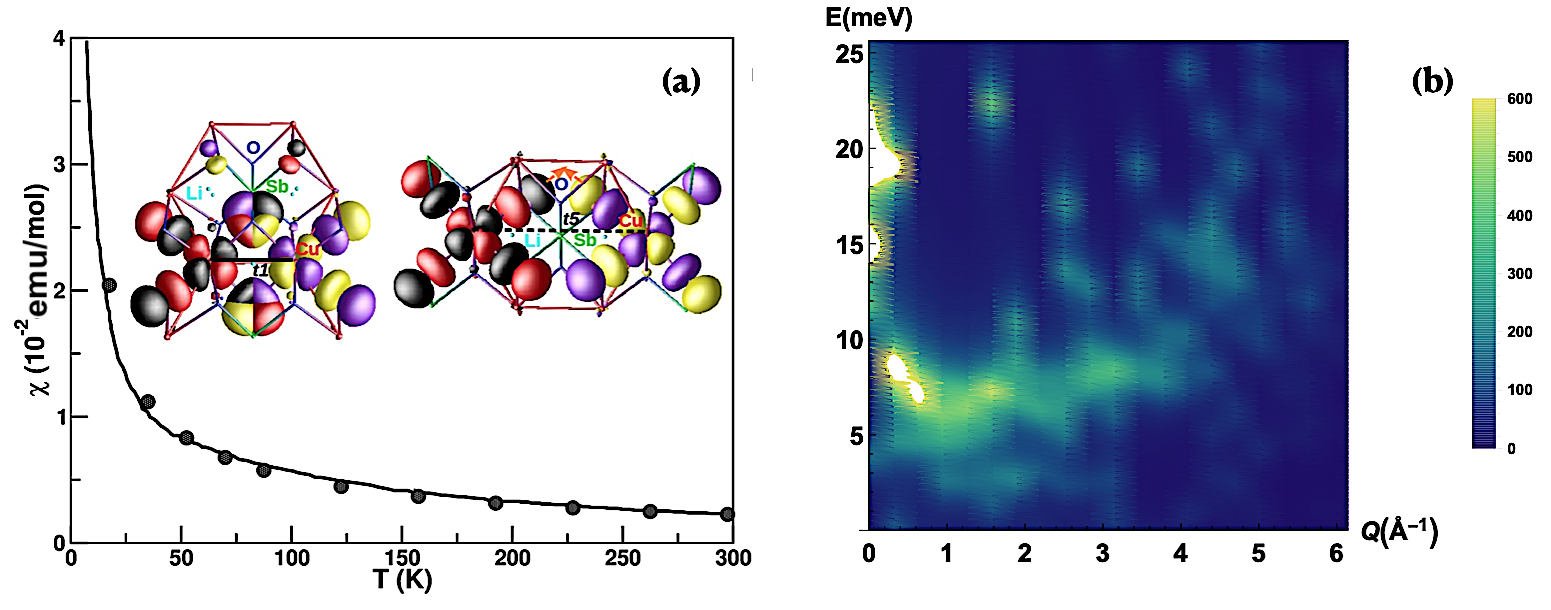}
\caption{ (a) Calculated spin susceptibility (circles) of the fragmented FM-AFM S = 1/2 chain model of Li$_3$Cu$_2$SbO$_6$ compound in comparison to the experimental data. The inset shows the plot of effective Cu x$^{2}$-y$^{2}$ Wannier functions placed at two neighboring Cu sites connected by t$_1$ hopping (left) and t$_5$ hopping (right). See text for details. (b) The calculated dynamical structure factor using the Exact diagonalization (ED) method.}
\end{figure}

\noindent {\bf Theoretical Calculations.} In order to achieve a theoretical understanding of the electronic and magnetic behavior of Li$_3$Cu$_2$SbO$_6$ we have carried out first-principles density functional theory (DFT) calculations~\cite{dft}, and solution of the DFT derived spin Hamiltonian by quantum Monte Carlo and exact diagonalization. The non spin-polarized band structure, assuming a perfectly ordered compound with no intermixing between Li and Cu sites, shows two Cu x$^2$-y$^2$ bands at the Fermi level strongly admixed with O $p$ and Sb states, corresponding to two Cu$^{2+}$, $d^9$ ions in the monoclinic $C2/m$ unit cell. The magnetic moments at Cu, O, and Sb in a spin polarized calculation turned out to be 0.49 $\mu_B$, 0.09 $\mu_B$, and 0.02 $\mu_B$, respectively. Since Li atoms are mobile there is disorder in the system, with some of the Li atoms replacing the Cu atoms, as found experimentally which is mimicked by considering a 1 $\times$ 16 $\times$ 1 supercell, resulting into 32 Cu sites in the cell, out of which some of the Cu and Li positions are interchanged, amounting to 6.25 $\%$ disorder. Starting from such a structure, to derive the underlying spin model, we use a muffin-tin orbital (MTO) based NMTO-downfolding calculation\cite{nmto} in which effective Cu x$^2$-y$^2$ Wannier functions are constructed by integrating out all the degrees of freedom other than Cu x$^2$-y$^2$, defining a low energy Hamiltonian. The real-space representation of the low energy Hamiltonian, show inter-layer Cu-Cu interactions to be negligibly small, with two dominant intra-layer Cu-Cu hoppings, t$_1$ and t$_5$, one connecting nearest neighbor edge-sharing Cu atoms, and another connecting Cu atoms through super-exchange paths involving O-Sb-O. The effective Cu x$^2$-y$^2$ Wannier functions are shaped according to x$^2$-y$^2$ symmetry, while the tails are shaped according to integrated O $p$ symmetries, due to large admixture between Cu x$^2$-y$^2$ and O $p$ (see inset in Figure 4(a)). For the nearest neighbor interaction (t$_1$) the O $p$ like tails of two neighboring Wannier functions are orthogonal to each other, for the interaction through super-exchange paths involving O-Sb-O (t$_5$) they point towards each other (marked with an arrow) in the figure. This makes the t$_5$ hopping 3.5 times stronger than t$_1$, although the Cu-Cu distance is 2.93 $\AA$ for t$_1$ and 5.79 $\AA$ for t$_5$. The edge-shared nearest-neighbor Cu atoms are connected through Cu-O-Cu bond angles of 89 and 86 degrees, which gives rise to the possibility of ferromagnetic exchanges in the system. Total energy calculations in-plane wave basis\cite{vasp,vasp2} of different magnetic configurations of Cu spins, and subsequent mapping to Heisenberg model, show edge-shared nearest-neighbor Cu-Cu magnetic interaction $J_1$ corresponding to hopping interaction t$_1$ to be ferromagnetic with a value - 21 meV while the long-ranged Cu-Cu magnetic interaction $J_2$ through Sb, corresponding to hopping interaction t$_5$ to be antiferromagnetic with a value 31 meV. Our ab-initio results thus predict a S = 1/2 FM-AFM alternating chain model with $J_{AFM}$ $\ge$ $J_{FM}$, in contrast to conclusions drawn in the previous work based on fitting of susceptibility data, suggesting $J_{AFM}$ $\le$ $J_{FM}$.\cite{koo2016static}

\noindent Based on first-principles input, we next consider a system of fragmented FM-AFM ($J_{FM}$ = - 21 meV and $J_{AFM}$ = 31 meV) S = 1/2 chains with 200 sites, and random disorder $\approx$ 6.25$\%$, given by the Hamiltonian, $\mathcal{H}=\sum_i^{N/2}[J_{AFM}S_{2i-1}.S_{2i}+J_{FM}S_{2i}.S_{2i+1}]-h\sum_i^NS^z_i$, where $i$ denote sites occupied by Cu atoms and $h$ is applied magnetic field, which was taken to be zero in our calculation. The impurity sites are chosen randomly. They host nonmagnetic Li atoms, are obtained by replacing Cu atoms of the pristine compound. The obtained results are averaged over 50 random configurations. In Stochastic Series Expansion implementation of Quantum Monte Carlo (SSE-QMC)\cite{qmc,qmc2} we measure the spin susceptibility as $\chi_{th} = \beta J < S_z^2- < S_z >^2 >$, where $\beta$ = 1/$k_BT$, which can be related to the experimentally measured molar susceptibility as 0.375~S(S+1)$g^2\frac{\chi_{th}}{T_J}$, where $T_J$ is the temperature corresponding to dominant magnetic exchange $J_{AFM}$. The comparison between the calculated and measured susceptibility is shown in Figure 4(a). Good matching between the two justifies the goodness of the ab-initio derived spin model.

\noindent Following the successful description of the experimentally measured susceptibility results, we attempt to calculate the inelastic neutron scattering response, which measures the magnonic excitations in a quantum spin system. Theoretically, the INS amplitude can be obtained from the calculation of frequency and momentum dependent dynamical structure factor~\cite{klauser2011spin} given by, $A^l(Q,\omega)=\sum_n|<\psi_n|S^l_Q|\psi_0>|^2\delta(\omega-(E_n-E_0))\nonumber =-\frac{Im[G(Q,\omega)]}{\pi}$. Here $l\sim x,y,z$, $|\psi_n>$ is $n$-th eigenvector of the Hamiltonian having energy eigenvalue $E_n$. $G(Q,\omega)$ denotes the dynamical correlation function or Green's function, which can be written (for $l$ = $z$) in terms of continued fraction~\cite{con} as, $G(Q,\omega)=\frac{<\psi_0|{S^z_Q}^\dagger S^z_Q|\psi_0>}{\omega+i\eta-a_0-\frac{b_1^2}{\omega+i\eta-a_1-\frac{b_2^2}{\omega+i\eta-...}}}$, where $S^z_Q$ is the Fourier transform of spin-$z$ operator $S^z_r$ and is given by $S^z_Q=\frac{1}{\sqrt{N}}\sum_rexp[iQ.r]S^z_r$. The continued fraction can be solved iteratively first by defining $|f_0>=S^z_Q|\psi_0>$ and obtaining the orthogonal states, $|f_{n+1}>=(\mathcal{H}-a_n)|f_n>-b_n^2|f_{n-1}>$, with $a_n=<f_n|\mathcal{H}|f_n>/<f_n|f_n>$, $b_{n+1}^2=<f_{n+1}|f_{n+1}>/<f_n|f_n>$ and $b_0=0$. The result obtained by averaging over results from 50 random configurations is shown in Figure 4(b). The overall features resemble well with measured INS data. In particular, large structure factor values are obtained around an energy $\approx$ 8-10 meV, as also seen in experimental data at 7 K. Similar to INS result, the calculated spectrum shows a low energy peak near $Q$ = 0 which indicates almost parallel preferential spin orientations among the nearest neighbor S = 1/2 Cu$^{2+}$ ions. This causes a FM like behavior even though the strongest interaction is antiferromagnetic. Due to the vacancy created in the spin-lattice as a result of the random replacement of the S = 1/2 Cu$^{2+}$ ions by nonmagnetic Li, an overall nonmagnetic spectral behavior dominates. \\

\noindent {\bf Summary.} The important finding is that Li$_3$Cu$_2$SbO$_6$ does not order magnetically down to 50 mK even though the system has a significant value of next nearest neighbour (NNN) AFM exchange interaction of $J_{AFM}$ = 31 meV, compared to NN FM $J_{FM}$ = -21 meV, confirmed through our bulk as well as microscopic measurements. The plateau-like behavior in $\lambda_{ZF}(T)$ and $\lambda_{LF}(H)$ might indicate the development of disorder state at low temperature due to the competing exchange interactions of $J_{FM}$ and $J_{AFM}$ arising from nearest neighbors and next-nearest neighbors. Finding real QSL materials is a rare phenomenon as there are only a few candidates reported so far, such as in pyrochlore lattice, kagome lattice, and organic charge-transfer salts with the frustrated triangular geometry, etc. Honeycomb 3$d$ layered oxides with a $d^9$ or $d^7$ quasi-two-dimensional lattice A$_3$A$^\prime_2$BO$_6$ (A = Li, Na; A$^\prime$ = Co, Ni; B = Sb, Te) could be potential candidates of spin liquid like state as there are numbers of materials with various stacking orders of the honeycomb slabs that are seen, but these systems are not studied yet in detail. \\

\noindent {\bf Acknowledgements.} We are grateful to Bella Lake for an interesting discussion. We would like to acknowledge financial support from the Department of Science and Technology, India (SR/NM/Z-07/2015) for the financial support and Jawaharlal Nehru Centre for Advanced Scientific Research (JNCASR) for managing the project. AB would like to acknowledge the Department of Science and Technology (DST) India for Inspire Faculty Research Grant (DST/INSPIRE/04/2015/000169), and the UK-India Newton grant for funding support. TKB would like to thank the Department of Science and Technology (DST), Government of India, for providing financial assistance in the form of DST-INSPIRE fellowship (IF160418). DTA would like to thank the Royal Society of London for the UK-China Newton funding and the Japan Society for the Promotion of Science for an invitation fellowship. SK acknowledges the Institute of Physics for providing computational facility and DST-SERB, Government of India for financial assistance (SRG/2019/002143). AMS thanks the SA-NRF (93549) and the UJ URC/FRC for generous financial assistance. TSD acknowledges J.C.Bose 
Fellowship (grant no. JCB/2020/000004) for research support. All data presented in this work are publicly available with identifier (DOI): https://doi.org/10.5286/ISIS.E.RB1968007, https://doi.org/10.5286/ISIS.E.RB1868023

\bibliographystyle{apsrev4-1}
\bibliography{refs}

\newpage
\newpage

\section{Supplemental Material}

\begin{figure}[b]
\centering
\includegraphics[width=0.75\linewidth]{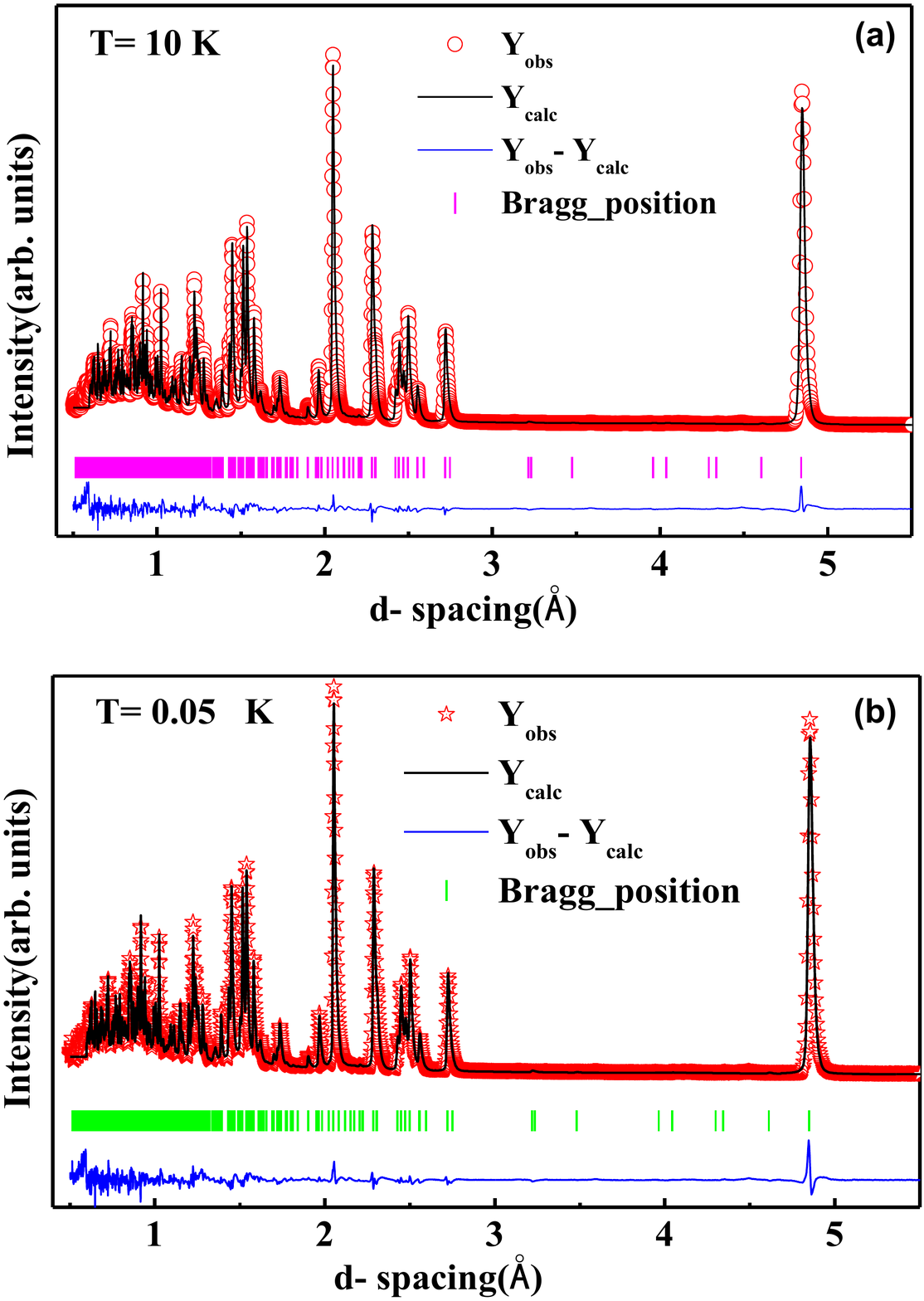}
\caption{Neutron diffraction data of Li$_3$Cu$_2$SbO$_6$ at 10 K (a) and 50 mK (b) with $d$-spacing.}
\label{zfmusr}
\end{figure}

\begin{figure}[b]
\centering
\includegraphics[scale=0.3]{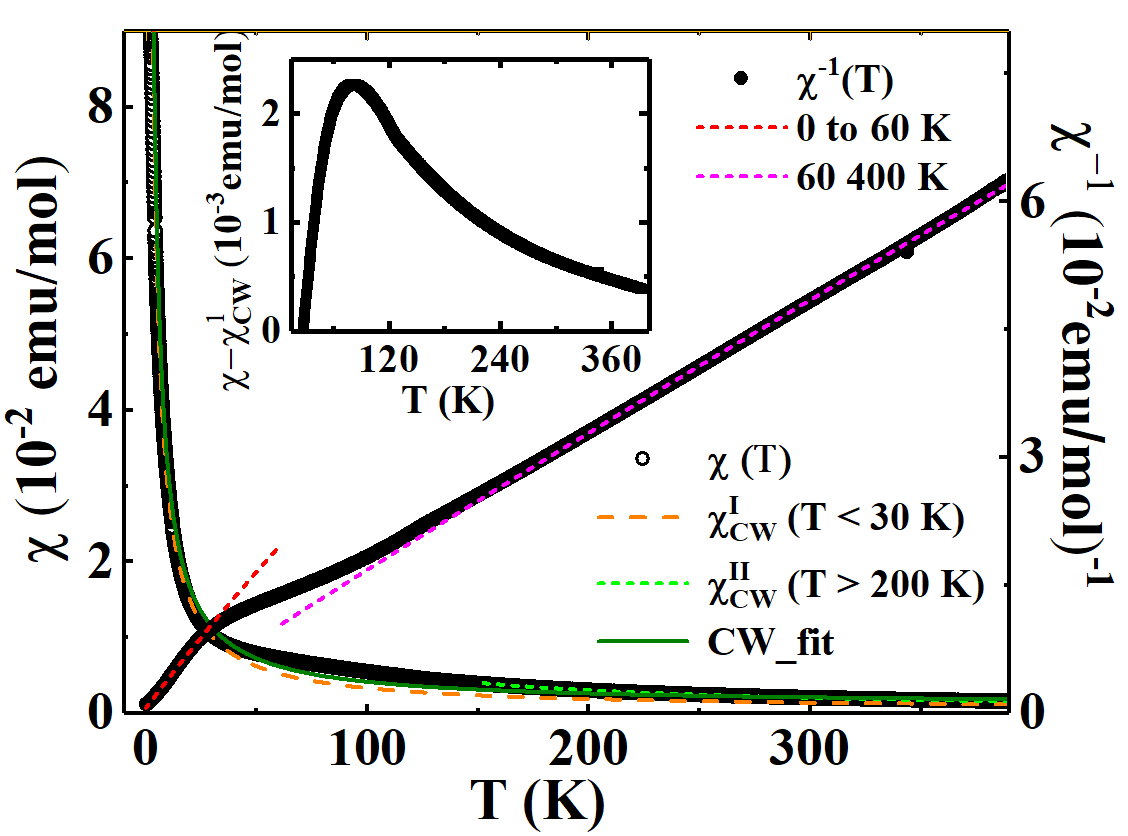}
\includegraphics[scale=0.33]{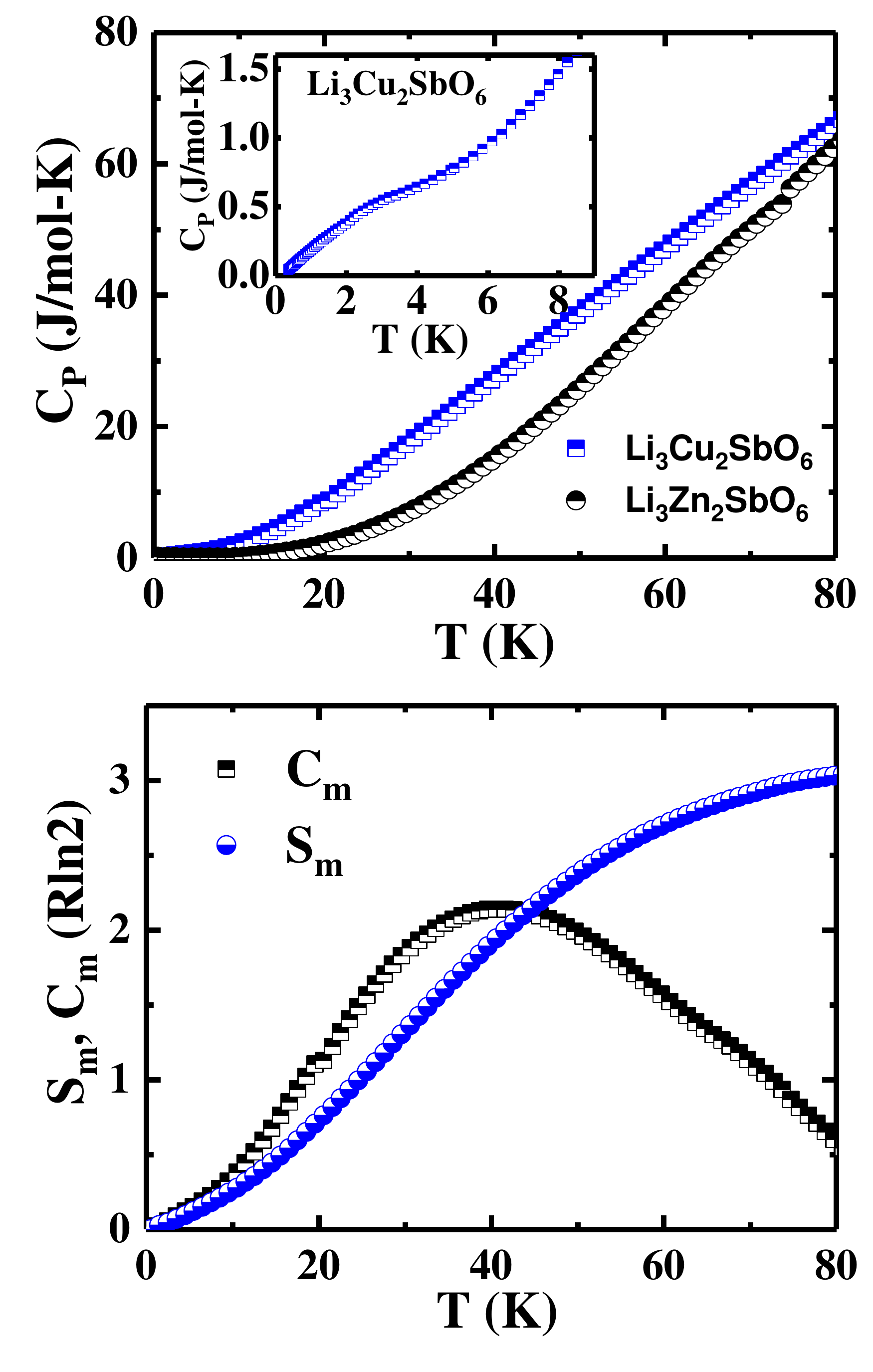}
\caption{(a) Temperature dependence of dc susceptibility and inverse of $\chi(T)$. Inset shows the bulk magnetic contribution of $\chi(T)$ after subtracting the low temperature CW-term. (b) Heat capacity as a function of temperature for Li$_3$Cu$_2$SbO$_6$ and Li$_3$Zn$_2$SbO$_6$. (c) Magnetic contribution of heat capacity and entropy plotted with temperature.}
\label{heatcapacity}
\end{figure}

\noindent {\bf Crystal Structure: Absence of long-range magnetic ordering.} To examine the crystal and magnetic structure of Li$_3$Cu$_2$SbO$_6$, high-resolution neutron powder diffraction data were collected on the WISH time-of-flight diffractometer at the ISIS Pulsed Neutron and Muon Source of Rutherford Appleton Laboratory, U.K. The sample ($\sim $ 6 g) was loaded into a cylindrical 6 mm vanadium can and measured on warming between 50 mK and 10 K using an Oxford Instruments cryostat. To check the crystal and magnetic structure of  Li$_3$Cu$_2$SbO$_6$, Rietveld refinements were performed using the FullProf program \cite{fullprof}. The high-temperature pattern at $T$ = 10 K was satisfactorily fitted using distorted honeycomb structure with the space group $C2/m$ ~\cite{koo2016static} for Li$_3$Cu$_2$SbO$_6$. Figure 1(a-b) shows the neutron diffraction data collected at 10 K and 50 mK. There is no difference between these two NPD data and no new peak detected down to 50 mK, which suggests the absence of long-range ordering in this material. Refinement of the nuclear peaks at 10 K and 50 mK confirms that it remains single-phase, and it crystallizes with the monoclinic $C2/m$ space group. So there is no structural phase transition in Li$_3$Cu$_2$SbO$_6$ down to 50 mK. The crystallographic parameters have been determined from bank 3 and summarized in Table S1 and S2.  The determined lattice parameters $a$ = 5.4707(5) \AA, $b$ = 8.6961(8) \AA, $c$ = 5.3645(3) \AA, and $\beta$ = 115.2021(1)$^{\circ}$ are in good agreement with Ref. [2]~\cite{koo2016static}.  The Wyckoff 4h (0.5, 0.362, 0.5) site was fully occupied by Li atom, whereas the 2c (0.5, 0.5, 0.50) Li atomic site was mixed with 21\% Cu atom. The site mixing of Li and Cu ion of Li$_3$Cu$_2$SbO$_6$ in $C2/m$ symmetric configuration occurs because of the nearly equal ionic radius of Li and Cu ion (0.76 and 0.73 I.R/{\AA}, respectively). For the ion exchange preparation of Li$_3$Cu$_2$SbO$_6$, Koo et al.~\cite{koo2016static} shows that there are 0.34 Cu$^{2+}$ ion in Li- sites. 

\noindent The refined ND data shows that there is a 7$\%$ site mixing between non-magnetic Li and Cu$^{2+}$ ions at the 4$g$ atomic site. Cu$^{2+}$-ions are separated by non-magnetic Sb$^{5+}$ ions and the dominating exchange interactions include a Cu-O-Cu superexchange mechanism with an ideal bond geometry of $\approx$ 90$^\circ$. This suggests putative frustrated magnetism and concurrence to the spin fluctuation regime. In the crystal structure of Li$_3$Cu$_2$SbO$_6$ with S = 1/2 Cu$^{2+}$ there are Cu-chains along the $b$-axis with nearest neighbour (NN) Cu-Cu distances shorter than $d1$ = 2.8409(7) \AA, while a longer distance for NNN $d2$ = 5.8553(1) \AA. The interchain distance in the ab-plane is $d3$ = 3.1231(2) \AA, while the 2D Cu$^{2+}$ planes are separated along the $c$-axis with a distance $d4$ = 5.3645(3) \AA. Rietveld refinement results are given in the Supplementary Tables 1 and 2.\\ 

\noindent {\bf Heat Capacity: Short-range ordering.} Results of specific heat ($C_P$) measurements down to 50 mK at zero applied magnetic field is shown in Figure 2(b). The total specific heat is expressed as a sum of magnetic and lattice contributions: $C_P = C_{mag} + C_{lat}$. The lattice contribution is subtracted by the heat capacity of the non-magnetic compound Li$_3$Zn$_2$SbO$_6$. The broad peak of $C_{mag}$ at 40 K as shown in Figure 2(c), is due to the opening of a spin gap from the Cu-Cu dimer formation. Similar features of $C_{mag}$, both in zero field and in field, have been seen in other spin liquid candidates~\cite{Balz2016}. In the case of A$_3$Ni$_2$SbO$_6$ (A = Li, Na)~\cite{Zvereva2015Li/Na3Ni2SbO6} the magnetic entropy released below $T_N$ removes less than 40\% of the saturation value, which indicates the presence of significant short-range correlations far above $T_N$.

\begin{figure}[b]
\centering
\includegraphics[width=\linewidth]{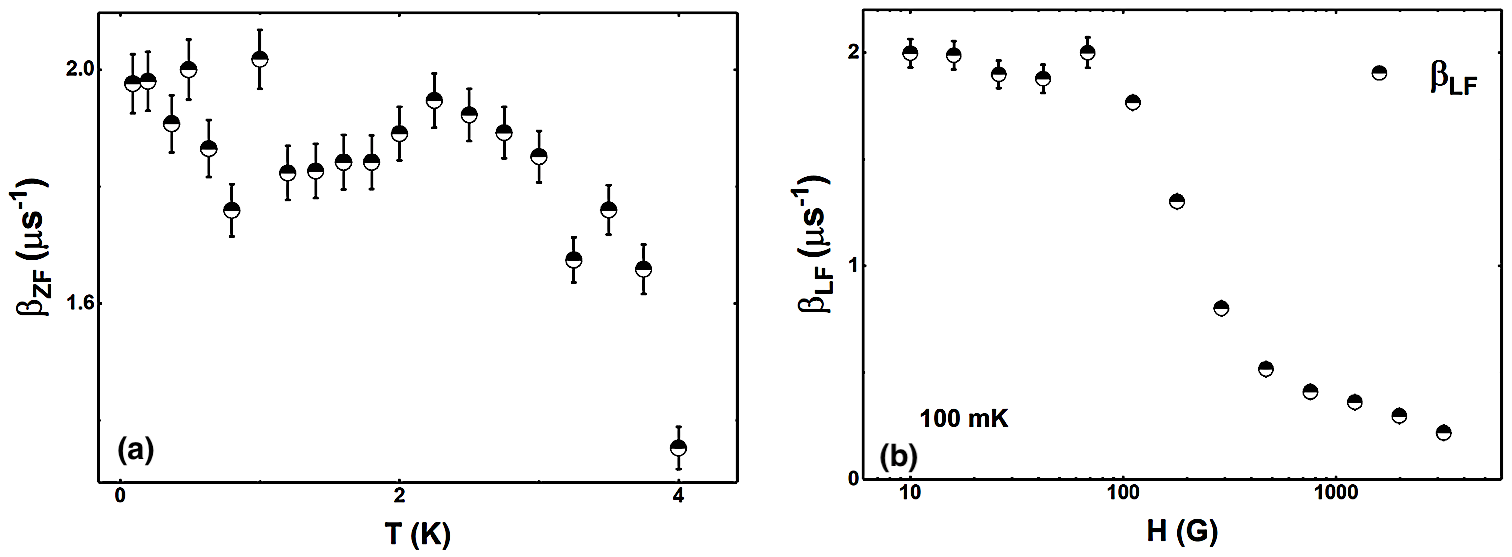}
\caption{(a) and (b) represent the temperature and field dependence of stretched exponent $\beta_{ZF}$ of Li$_3$Cu$_2$SbO$_6$.}
\label{zfmusr}
\end{figure}

\begin{figure}[t]
\includegraphics[width=\linewidth]{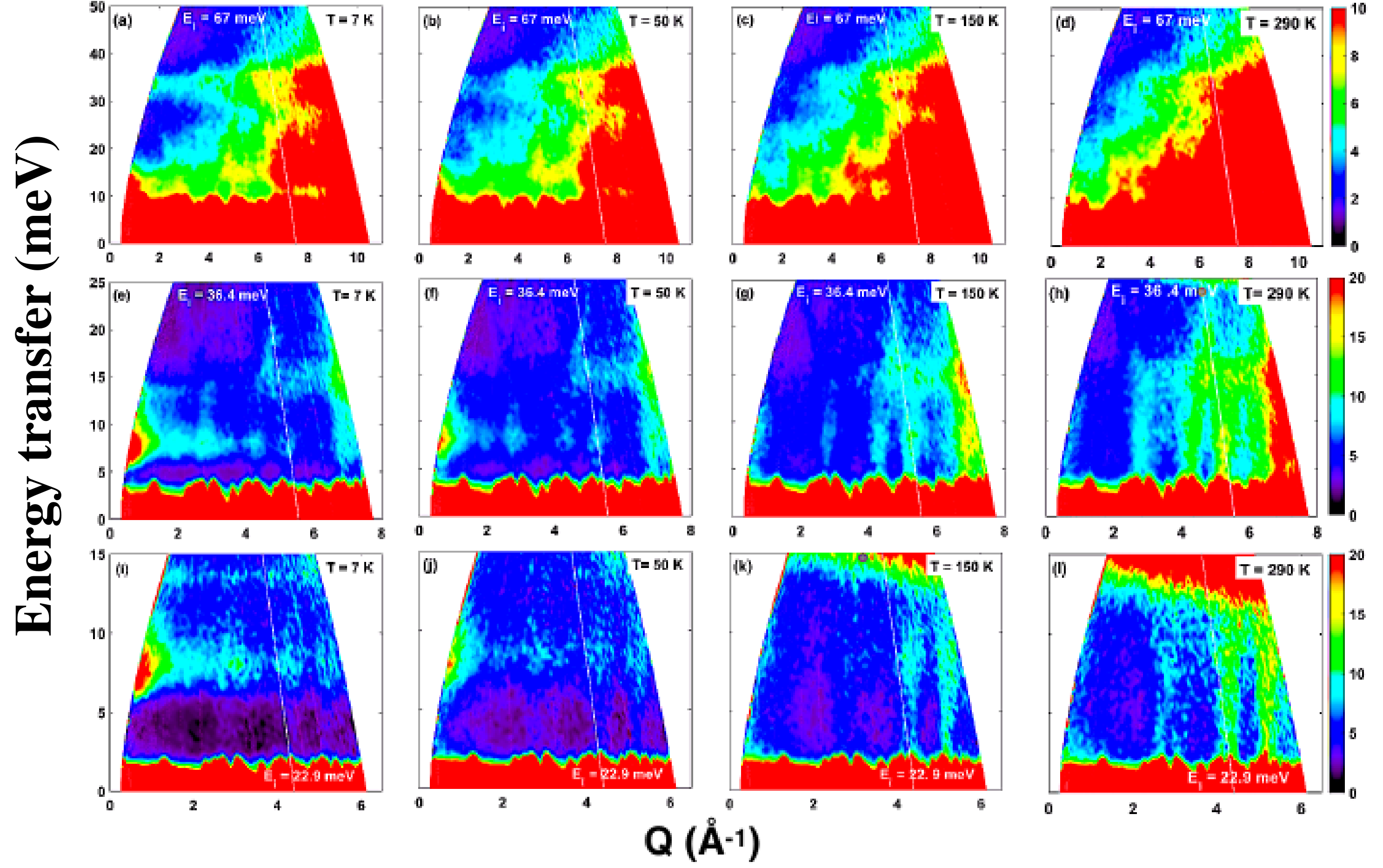}
\caption{Colour coded intensity maps of the scattering intensity plotted as energy transfer versus momentum transfer from the powder sample of Li$_3$Cu$_2$SbO$_6$ (a-d) with E$_i$ = 67 meV, (e-h) E$_i$ =36.4 meV and (i-l) E$_i$ = 22.9 meV at 7 K, 50 K, 150 K, and 290 K.}
\label{INS}
\end{figure}

\begin{figure}[t]
\includegraphics[width=\linewidth]{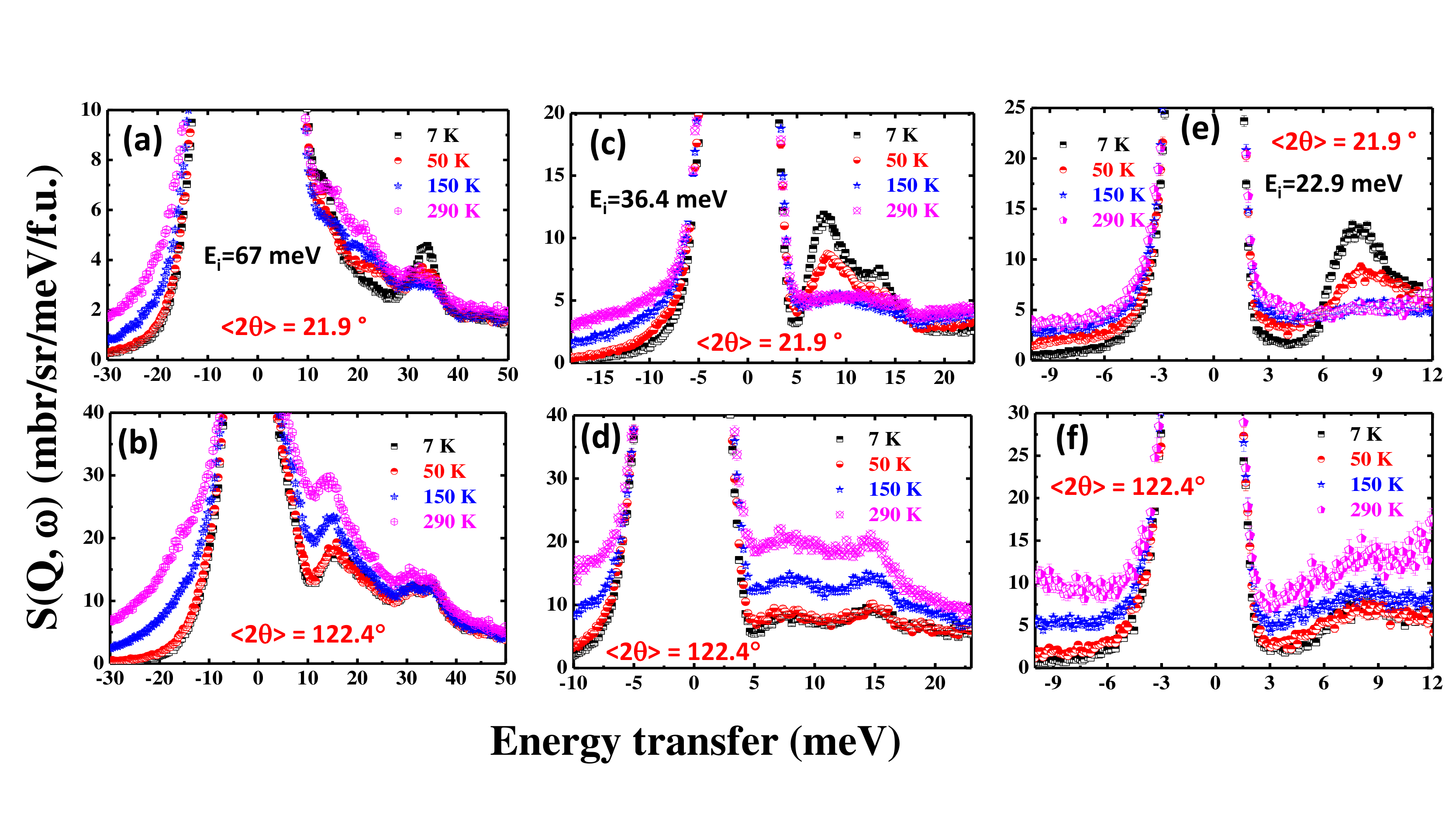}
\caption{The scattering angle integrated one-dimensional cuts, Intensity vs. Energy transfer at various temperatures (top raw: a, c, e) from low angles where the magnetic scattering is dominated and (bottom raw: b, d, f) from high angles where the phonon scattering is dominated. The cuts were made in the scattering angles (rather than $Q$) to avoid spurious scattering at low scattering angles coming from the divergence of the direct beam.}
\label{INS}
\end{figure}

\begin{figure}[t]
\includegraphics[width=\linewidth]{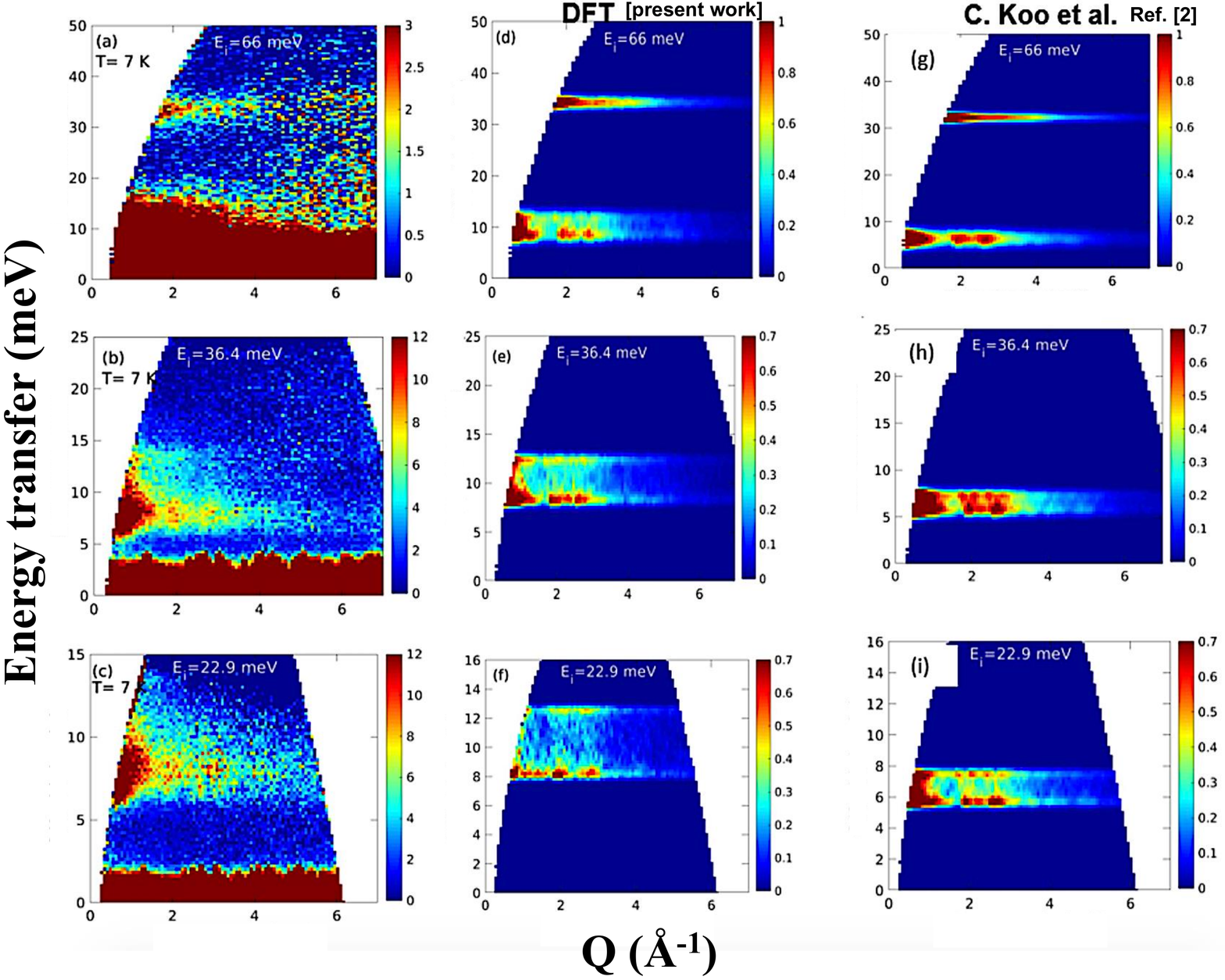}
\caption{(Left) The estimated magnetic scattering, after phonon subtracted, for neutron incident energy of $E_i$ = 66 meV (a), $E_i$ = 36.4 meV (b)  and $E_i$ = 22.9 meV (c) at 7 K  for Li$_3$Cu$_2$SbO$_6$. (Middle, d-f) Simulated magnetic excitations using the linear spin wave theory (LSWT) using the exchange parameters $J_{1}$ = $J_{FM}$ = - 21 meV and $J_2$ = $J_{AFM}$ =  31 meV,  estimated using our DFT calculation (see text in the main paper) and a single ion anisotropy D$_y$ = -2 meV along the b-axis.  (Right, g-i) Simulated magnetic excitations using the exchange parameters $J_1$ = 24.45 (FM) and $J_2$ = -13.79 meV (AFM) (from Koo et al, [2]~\cite{koo2016static}) and D$_y$ = -2 meV. We have used spin-W program to simulate the magnetic excitations spectra using LSWT (4). The linewidths of the observed excitations (4.1$\pm$0.2 meV for 33 meV peak, 4.8$\pm$0.2  meV for 13 meV peak and 3.7$\pm$0.2  meV for 8 meV peak) are higher than the instrument resolution.}
\label{INS}
\end{figure}

\noindent {\bf Muon spin relaxation spectroscopy.} To probe the potential magnetic order or spin freezing, we have performed muon spin relaxation spectroscopy. Experimentally, dynamic spin fluctuations and static magnetic order can be detected by $\mu$SR due to the extreme sensitivity of the muon to small magnetic moments. The $\mu$SR data was collected at the ISIS pulsed muon facility using the Oxford dilution refrigerator between 80 mK and 6 K. Additional data between 2 and 180 K were collected by transferring the sample to a $^4$He cryostat. Data taken in these two cryostats agree very well in the overlapping temperature region. Detailed analysis of the observed relaxation function and its longitudinal field dependence reveals an unconventional time evolution of the local fields below 4.0 K, which may be related to the formation of the quantum spin liquid like state of Li$_3$Cu$_2$SbO$_6$. Figure 3(a-b) represent the temperature and field dependence of stretched exponent $\beta_{ZF}$ of Li$_3$Cu$_2$SbO$_6$.

\noindent {\bf Inelastic neutron scattering excitations.}  Now we compare the energies of the INS excitations observed on our powder sample of Li$_3$Cu$_2$SbO$_6$, with that observed in the single crystals of Na$_3$Cu$_2$SbO$_6$ having same crystal structure~\cite{miura2008magnetic} and very similar susceptibility behaviour (maximum near 85 K)  as seen in the inset of  Figure 2(a)  for Li$_3$Cu$_2$SbO$_6$. INS study of Na$_3$Cu$_2$SbO$_6$ exhibits a spin gap of 8.9 meV at 10 K, and the gap increases to 10 meV at 200 K. At 10 K, the excitations exhibit dispersion with a zone center (ZC) energy of 8.9 meV and zone boundary (ZB) energy of ~15 meV. In Na$_3$Cu$_2$SbO$_6$ the exchange interactions between the neighboring spins with the shorter and longer spacings correspond to ferromagnetic ($J_{FM}$ = -12.5 meV) and antiferromagnetic ($J_{AFM}$ = 13.9 meV) interactions, respectively~\cite{miura2008magnetic}. Considering the rather large Jahn-Teller distortion acting on the Cu-2p spins in the $x^2-y^2$ orbit (with local coordinates) in Na$_3$Cu$_2$SbO$_6$, the interchain coupling was neglected, and the observed excitations were explained using an alternating chain model along the $b$-axis with $J_{FM}$ and $J_{AFM}$~\cite{miura2008magnetic}. Further, from the microscopic measurement $J_{FM}$ = -18.03 meV and $J_{AFM}$ = 14.2 meV were obtained for Na$_3$Cu$_2$SbO$_6$~\cite{miura2008magnetic}. It is to be noted that there are no reports of high energy excitations in Na$_3$Cu$_2$SbO$_6$ as we have observed 33 meV excitations on Li$_3$Cu$_2$SbO$_6$. We attribute the two lower energy excitations 7.7 meV and 13.5 meV arising from the zone center and zone boundary of the low energy model due to the powder averaging effect in our powder sample of Li$_3$Cu$_2$SbO$_6$. The detail on this is given in the Supplementary information where we have calculated the excitations using the linear spin-wave theory and using the spin-W program~\cite{toth2015linear}, which reproduce observed three excitations very well as shown in Supplementary Figure 6.  

\noindent The inelastic neutron scattering measurements were performed on a powder sample of Li$_3$Cu$_2$SbO$_6$ using the high neutron flux spectrometer MERLIN at the ISIS Neutron scattering facility UK. Powder sample of 9 g was loaded in an annular aluminum can with a diameter of 40 mm and height of 40 mm, which was inserted into a closed-cycle refrigerator under He-exchange gas that operated between 5 K and 300 K. The measurements were performed with an incident energies $E_i$ = 160 meV with Gd-chopper frequency of 500 Hz (this also gave data of $E_i$ = 67, 36.4 and 22.9 meV) and 100 meV with Gd-chopper frequency of 350 Hz (this also gave data with $E_i$ = 38 and 20 meV) in multi-$E_i$ model at various temperatures between 5 K and 300 K.  We also measured a standard vanadium sample at the same set of incident energies to convert the measured sample's scattering intensities into normalized units of cross-section, (mbr/sr/meV/f.u.), where f.u. stands for formula unit of Li$_3$Cu$_2$SbO$_6$.\\

\noindent Considering that we have estimated exchange parameters from our DFT calculations, which reveal stronger AFM exchange, while the exchange parameters reported by Koo et al.~\cite{koo2016static}, reveal stronger FM exchange, it is important to find out which set of parameters are reliable. To find out which set of  parameters reproduce the observed magnetic excitations in Li$_3$Cu$_2$SbO$_6$, we have simulated the magnetic excitations of Li$_3$Cu$_2$SbO$_6$ using the linear spin wave theory (LSWT). We have used spin-W program to simulate the magnetic excitations~\cite{toth2015linear}. The energy dependent instrument resolution was included in our simulation. The intensity of simulated magnetic excitations was also corrected by the Cu$^{2+}$ magnetic form factor. \\

\begin{table}
\centering
\caption{Rietveld refined lattice parameters and bond angle of Li$_3$Cu$_2$SbO$_6$ at 10 K and 50 mK.}
 \begin{tabular}{c c c }
 \hline 
 Lattice Parameter& 10 K& 50 mK \\  
 \hline
 a (\AA) & 5.4707(5) & 5.4692(9)\\

 b (\AA) & 8.6961(8)	& 8.6936(7)\\

 c (\AA) & 5.3645(3)	& 5.3646(8)\\

 $\beta$ (deg) & 115.2020(1)	& 115.2047(2)\\

 Volume (\AA$^3$) & 230.9199 & 230.793\\

 Bragg R-factor & 5.10 & 5.33\\
\hline
\end{tabular}
 \label{tab:my_label}
\end{table}
\begin{table}[ht]
  \centering
  \caption{Atomic coordinates and occupancy parameters of Li$_3$Cu$_2$SbO$_6$.}
  \label{tab:my_label2}
  \begin{tabular}{c c c c c c}
  \hline 
  Atom & x & y & z & Occupancy & Site\\  
  \hline 
  Li1	& 0.5000(0) & 0.3626(1) & 0.5000(0) & 1.00(0) & 4h\\
   
  Li2 & 0.5000(0) & 0.5000(0) & 0.5000(0) & 0.784(1) & 2c\\
   
  Cu2 & 0.5000(0) & 0.5000(0) & 0.5000(0) & 0.215(1) & 2c\\
   
  Cu3	& 0.0000(0) & 0.3366(6) & 0.0000(0) & 0.928(2) & 4g\\
   
  Li3	& 0.0000(0) & 0.3366(6) & 0.0000(0) & 0.072(2) & 4g\\
  
  Sb1 & 0.0000(0) & 0.0000(0) & 0.0000(0) & 1.000(0) & 2a\\
   
  O1 & 0.2436(2) & 0.0000(0) & 0.7978(1) & 1.000(0) & 4i\\
   
  O2 & 0.2281(1) & 0.1700(2) & 0.2419(5) & 1.000(0) & 8j\\
   
  \hline
\end{tabular}
\end{table}

\noindent In crystal structure of  Li$_3$Cu$_2$SbO$_6$ with Cu$^{2+}$ S = 1/2 there are Cu-chains along the b-axis with NN Cu-Cu distances shorter $d$1 = 2.8409(7) \AA~($J_1 = J_{FM}$) , while longer distance for NNN $d$2 = 5.8553(1) \AA~($J_2 = J_{AFM}$).  The interchain distance in the ab-plane is $d$3 = 3.1231(2) \AA~(J$_3$), while the 2D Cu-plane are separated along the $c$-axis with distance $d$4 = 5.3645(3) \AA. As DFT calculation suggested that important exchange parameters are along the b-axis $J_1$ and $J_2$, we have used only these two parameters in our simulation of magnetic excitations in Li$_3$Cu$_2$SbO$_6$.  The magnetic ground state of Li$_3$Cu$_2$SbO$_6$ was assumed same as in the isostructural Na$_3$Co$_2$SbO$_6$  with a propagation vector $k$ = [1/2 1/2 0] ~\cite{JQYan2019structure}. First, we calculated the magnetic excitations of Li$_3$Cu$_2$SbO$_6$ without any anisotropy, which gave gapless excitations. Considering that observed magnetic excitations are gapped (8 meV), we added single ion anisotropy term in the Hamiltonian.  We simulated the magnetic excitations with single ion anisotropy, planner and axial anisotropies. We found the best agreement between the experimental and calculated spectra was obtained with the axial anisotropy along the b-axis (i.e. D$_y$ = -2 meV). Figure 6 (middle and right) shows the simulated magnetic excitations (powder average scattering) for two sets of exchange parameters, from our DFT and from Koo {\it et. al}~\cite{koo2016static}. It is clear that the simulation gives two excitations at low energy (between 5 to 13 meV) and one excitation at high energy (33 meV). The low energy two excitations arise from the powder average of zone center and zone boundary modes. Comparing the simulated spectra with the experimental results, it is clear that the DFT exchange parameters give the best agreement between theory and experimental results, which shows that $J_{FM} \le J_{AFM}$.  This result was also supported when we analyzed the energy integrated  $Q$-dependence of the magnetic intensity using a model independent analysis based on the first moment sum rule~\cite{MBSone2002frustration,Hohenberg1974magnetic,Hammar1998spingap}. This indicates that physics of  Li$_3$Cu$_2$SbO$_6$  can be explained well with the bond alternating S = 1/2 Cu$^{2+}$ chain with weak FM and strong AFM exchanges  rather than S = 1 Haldane chain (i.e. strong FM exchange). On the other hand in Na$_3$Cu$_2$SbO$_6$  the exchange interactions between the NN Cu-spins with the shorter distance and NNN Cu-spins with longer distance are ferromagnetic $J_{FM}$ = - 12.5 meV and antiferromagnetic $J_{AFM}$ = 13.9 meV, respectively~\cite{miura2008magnetic}.  The magnetic excitations of Na$_3$Cu$_2$SbO$_6$  are also gapped with an energy gap of 8.9 meV at 10 K and gap increases to 10 meV at 200 K~\cite{miura2008magnetic}. It is to be noted that there are no reports of high energy magnetic excitations in Na$_3$Cu$_2$SbO$_6$ as we have observed 33 meV excitations on Li$_3$Cu$_2$SbO$_6$. Hence, the further investigation of high energy excitations in Na$_3$Cu$_2$SbO$_6$ single crystals as well as single crystal study of Li$_3$Cu$_2$SbO$_6$ are important. 

\end{document}